# *In-situ* high-pressure Raman scattering studies in PbWO$_4$ up to 48 GPa

D. Errandonea[1,†], R. Lacomba-Perales[1], K. K. Mishra[2], and A. Polian[3,4]

[1] Departamento de Fisica Aplicada, Institut Universitari de Ciència dels Materials, Universitat de Valencia, c/ Doctor Moliner 50, E-46100 Burjassot, Valencia, Spain

[2] Condensed Matter Physics Division, Indira Gandhi Centre for Atomic Research, Kalpakkam, Tamil Nadu 603102, India

[3] IMPMC, Institut de Minéralogie, de Physique des Matériaux, et de Cosmochimie, Sorbonne Universités, UPMC Univ Paris 06, 4 Place Jussieu, F-75005, Paris, France

[4] CNRS, UMR 7590, F-75005, Paris, France

**Abstract:** The effect of pressure on the Raman spectrum of PbWO$_4$ has been investigated up to 48 GPa in a diamond-anvil cell using neon as pressure-transmitting medium. Changes are detected in the Raman spectrum at 6.8 GPa as a consequence of a structural phase transition from the tetragonal scheelite structure to the monoclinic PbWO$_4$-III structure. Two additional phase transitions are detected at 15.5 and 21.2 GPa to the previously unknown crystalline phases IV and V. The last one remains stable up to 43.3 GPa. At 47.7 GPa all Raman modes disappear, which could be caused by a pressure-induced amorphization. All structural changes are reversible, being the scheelite phase recovered at ambient pressure. However, the two most intense modes of the PbWO$_4$-III phase are still present after full decompression, indicating that this phase coexists as a minority metastable phase with the scheelite phase after pressure release. The wavenumber of the Raman modes and their pressure dependencies are reported for the four crystalline phases. The present reported results are compared with previous studies.

[†] Corresponding author; email: danel.errandonea@uv.es

1.  **Introduction**

The isostructural tungstates $CaWO_4$, $SrWO_4$, $BaWO_4$, and $PbWO_4$ are compounds which have been extensively studied during the last decade because of their interesting physical properties and their multiple applications [1, 2]. These materials can be growth as single crystals [3, 4] or be synthesized as nanoparticles [5, 6] to satisfy the requirements of different applications. The above mentioned orthotungstates crystallize in the tetragonal scheelite-type structure (space group $I4_1/a$, No. 88, Z = 4) at ambient conditions [7]. This crystal structure was first refined by Sillen and Nylander [8]. In the case of $PbWO_4$ (mineral name stolzite); Pb and W atoms are on special Wyckoff positions, Pb (W) on 4b (4a), and the oxygen atoms on a general 16f Wyckoff position [9]. A meticulous description of the scheelite structure of $PbWO_4$ can be found in the literature [10]. It basically consists of chains of a network of $WO_4$ tetrahedra and $PbO_8$ dodecahedra. Each tetrahedron shares corners with four vertices of the adjacent dodecahedra.

The high-pressure (HP) properties of scheelite-structure tungstates have been comprehensively studied for years. In the case of $PbWO_4$; Chang *et al.* in the early 70s already reported the existence of a monoclinic high-pressure high-temperature (HP-HT) polymorph ($PbWO_4$-III) which was quenchable to ambient conditions [11]. More than a decade later, Hazen *et al.* determined the unit-cell parameters of the scheelite-type structure at several pressures, being this phase reported to be stable to 6 GPa [12]. Simultaneously, Jayaraman *et al.* detected the occurrence of a phase transition beyond 4.5 GPa employing Raman spectroscopy [13]. Later on, in the 21[st] century, a combination of experimental and theoretical studies determined the transition pressure to be between 5.6 and 9 GPa [14 - 17]. The observed structural changes have important consequences in the optical and electronic properties of $PbWO_4$ [18, 19], including a band-gap collapse from 3.5 to 2.7 eV [18]. In addition, in two of the previous studies [14, 15], the occurrence of a second phase transition was reported to happen beyond 10 GPa. The phase transitions were found to take place from the tetragonal scheelite structure to the monoclinic fergusonite structure (space group *I2/a*, No. 15, Z = 4) followed by a denser monoclinic structure named $PbWO_4$-III (space group *P2$_1$/n*, No. 14, Z = 8) [14].

In contrast with experiments, *ab initio* calculations predicted that the phase transition should occur directly from the scheelite structure to the denser monoclinic structure (*P2$_1$/n*). The presence of the fergusonite structure in between both phases in the experiments was attributed to the presence of kinetic barriers [14, 15], but it could be also triggered by non-hydrostatic effects. In particular, non-hydrostatic conditions have been found recently to strongly influence the HP behavior of scheelite-type $BaWO_4$ and $CaWO_4$ [20 - 22]. As predicted by calculations, in $BaWO_4$ experiments found that under quasi-hydrostatic conditions the pressure-induced transition occurs from scheelite to a monoclinic structure with space group *P2$_1$/n*, being the fergusonite structure observed as an intermediate phase only in experiments carried out under non-hydrostatic conditions [20].

Clearly, in spite that $PbWO_4$ has been already studied under HP conditions, there are still fundamental questions concerning its structural properties under compression. These issues include not only the influence of non-hydrostatic effects on the HP behavior of $PbWO_4$, but also the existence of additional phase transitions and a possible pressure-driven amorphization, a common phenomenon in minerals related to scheelite [23], which in the case of $PbWO_4$ has been recently predicted to occur near 45 GPa [1]. To address these issues we report here, HP Raman experiments carried out using Ne as pressure-transmitting medium (PTM) to generate quasi-hydrostatic conditions. Our experiments extended the pressure range covered by previous Raman experiments from 17 GPa [15, 16] to 47.7 GPa. We found evidence of a direct phase transition from scheelite to the monoclinic $PbWO_4$-III structure at 6.8 GPa. No evidence of the existence of the previously reported fergusonite structure was found, confirming that as in $BaWO_4$ [20], non-hydrostatic effects influence the HP behavior of $PbWO_4$. We also will report evidence of two additional phase transitions which occur at 15.5 and 21.2 GPa. Finally, a possible amorphization of $PbWO_4$ was detected at 47.7 GPa. All the structural changes reported are partially reversible upon decompression. The reported results will be compared with previous studies.

2. **Experimental details**

The $PbWO_4$ samples used in the Raman experiments were obtained from a high-purity single crystal grown with the Czochralski method starting from raw powders having 5N purity [24, 25]. For the HP measurements a 5 μm thick plate was cleaved from the crystal and loaded in a 300 μm culet diamond-anvil cell (DAC) together with a few ruby spheres of about 2 μm in diameter. The sample was loaded into a chamber 100 μm in diameter and 40 μm thick drilled in an Inconel gasket. Ne was used as PTM [26], and pressure was determined by the ruby fluorescence technique [27]. The accuracy of the pressure determination was better than 0.1 GPa. The Raman experiments were performed at room temperature, in the backscattering geometry, using the 514.5 nm line of an $Ar^+$-ion laser. A laser power of less than 100 mW before the DAC was found to be appropriate for optimizing the signal–noise ratio. The temperature increase of the sample due to laser absorption was negligible because the laser energy is below the band gap of $PbWO_4$ [19]. A Mitutoyo 20× long working distance objective was employed for focusing the laser on the sample and for collecting the Raman spectra. The scattered light was analyzed with a Jobin-Yvon T64000 triple spectrometer equipped with a confocal microscope in combination with a low-background liquid-nitrogen-cooled multichannel charge-coupled device detector. In the experiments we used a 1200 grooves/mm grating and a 100 μm entrance slit. The spectrometer was calibrated with reference to the first-order Raman line of a single crystal of silicon and Ar plasma lines. The spectral resolution in our measurements was better than 1 cm$^{-1}$.

**3.    Results and discussion**

Fig. 1 shows room-temperature Raman spectra of PbWO$_4$ measured from ambient up to 13.1 GPa. Group-theoretical considerations, for the scheelite structure, lead to a total of thirteen zone-center Raman-active modes: $\Gamma = 3A_g + 5B_g + 5E_g$ [28]. In our experiments we detected all the thirteen expected Raman modes. They are indicated by ticks in Fig. 2 for the Raman spectrum measured at 0.2 GPa. Their wavenumbers (ω) at ambient pressure are given in Table 1 and compared with previous experiments [13, 15, 16]. The agreement is good. As pressure increases only the Raman modes of the scheelite polymorph (phase I) are detected up to 4.7 GPa. This is illustrated in Fig. 1 with the spectra measured at 2.7 and 4.7 GPa. In a subsequent compression step, at 6.8 GPa new Raman modes appear, but the scheelite modes are still dominant. This indicates the onset of a structural phase transition. At 6.9 GPa, the Raman modes of the HP phase become dominant and only the strongest Raman mode of scheelite remains detectable near 910 cm$^{-1}$. This mode corresponds to an internal symmetric stretching vibration of the WO$_4$ tetrahedron [29]. Under further compression the last surviving mode of scheelite disappears and only the modes of HP phase are detected up to 13.1 GPa (see Fig. 1).

Previously, a pressure induced transition was found in PbWO$_4$ at pressures ranging from 4.5 to 9 GPa. The transition pressure depends on the pressure-transmitting medium and the experimental technique [13 – 17]. In this case, under quasi-hydrostatic conditions, we found that the transition pressure is 6.8 GPa and that the low- and high-pressure phases coexist within a pressure range of 0.1 GPa. From the experimentally obtained mode positions, we determined the pressure dependence of the Raman modes of the low and high-pressure phases. Figs. 2 and 3 show the obtained pressure dependence of the different observed modes. There it can be seen that, for the scheelite phase and the high-pressure phase, the evolution of Raman modes with pressure can be described by a linear function. The obtained pressure coefficients ($d\omega/dP$) are given in Table 1. The agreement with previous experiments [13, 15, 16] carried out under less hydrostatic conditions is good. Previous *ab initio* calculations [15] also agree with our experiments. The only remarkable difference is that experiments found that the lowest frequency mode gradually softens under compression, a typical feature of scheelite-type oxides [30], while for this mode, theory predicts a small positive pressure coefficient. The effect of pressure on the Raman modes of scheelite-type $PbWO_4$ has been already extensively discussed in the literature [13, 15, 16]. Therefore, to avoid redundancies we will not discuss it into detail here, mentioning only that the discrepancy on the sign of the pressure coefficient of the lowest-frequency mode could be due to a small miscalculation of the effect of pressure in the inter-atomic distances. It should be also noted that calculations were carried out at 0 K within a harmonic approximation, which could also be a possible reason for this small discrepancy.

We will discuss now the Raman spectra of the HP phase. In addition to the scheelite polymorph, a monoclinic metastable polymorph has been reported for $PbWO_4$ at ambient conditions. This polymorph, named raspite (phase II) [31], cannot explain the large number of Raman modes we found in the HP phase. In total we detected up to forty-eight Raman modes, which are given at different pressures in Table 2, whereas the raspite structure only has eighteen Raman-active modes [32]. Consequently, it cannot be the HP phase detected for PbWO$_4$ in our experiments. Previous Raman experiments carried out using silicone oil as pressure-medium found the phase transition to occur at 6.2 GPa [15]. In this work, the HP phase was proposed to have the monoclinic crystal structure of the HP-HT PbWO$_4$-III phase. However, this phase was found to coexist with another monoclinic structure with a fergusonite-type structure (phase II') up to 14.6 GPa [15]. This conclusion was in agreement with the results obtained by XRD experiments performed using also silicone oil as pressure-medium [14]. In contrast, theory has predicted a direct transition from scheelite to PbWO$_4$-III. Our experiments are in agreement with this last result. All the forty-eight Raman modes detected for the HP phase in our measurements can be assigned to the PbWO$_4$-III structure. For this structure seventy-two Raman-active modes are predicted by group theory [15]. The increase of the number of Raman modes is due to the reduction of the crystal symmetry and the doubling of the size of the unit-cell [15]. We support our identification of the HP phase as PbWO$_4$-III with the fact that the forty-eight modes measured at 8.9 GPa agree well with the wavenumbers of the modes theoretically calculated at 8.7 GPa [15] as shown in Table 2. The fact that some of the calculated modes are nearly degenerated and that the Raman modes of the HP phase are broader than those of the low-pressure phase could explain why we observed only 48 of the 72 expected modes. We would like to note here, however, that we observed more Raman modes for the PbWO$_4$-III than in previous studies [15, 33]. The statement holds even when the comparison is made with ambient pressure studies [33] in which only thirty-one modes were measured. A comparison of the Raman modes here reported for PbWO$_4$-III with previous experiments is also shown in Table 2. The wavenumbers of many of the forty-eight modes we observed for the HP phase at 8.9 GPa agree well with those calculated from the results reported by Manjon *et al.* [15]. The agreement between experiments is also good for the frequencies measured around 13 GPa (see Table 2) and for the pressure coefficients as can be seen by comparing Table 2 from this work with Table 5 from Ref. 11. In addition, there is also a nice correlation when the comparison is made with the thirty-one modes measured for PbWO$_4$-III by Tan *et al.* [33]. In this case, the wavenumbers shown in Table 2 at 8.9 GPa are obtained from the wavenumbers and the pressure coefficients reported at ambient pressure, assuming a linear dependence of the phonon frequencies with pressure (as we observed in phase III).

On the other hand, in our experiments, we did not observe any of the modes previously assigned to fergusonite-type PbWO$_4$ [15]. The observation of a direct transition from scheelite PbWO$_4$ (phase I) to the PbWO$_4$-III structure, without going through the intermediate fergusonite-type structure, is in agreement with the hypothesis that fergusonite is observed only due the sluggishness and kinetic hindrance of the transition from scheelite to PbWO$_4$-III [15]. It is interesting to note here, that a similar phenomenon was observed in scheelite-type BaWO$_4$ [20] and SrMoO$_4$ [34]. In both compounds, the fergusonite-type structure appears as the post-scheelite phase when experiments were carried out using alcohol mixtures or silicone oil as pressure medium. However, they transform to structures isomorphic to PbWO$_4$-III when the experiments are carried out using Ne as pressure-medium, which is hydrostatic at the transition pressure (P < 10 GPa). This fact certifies that deviatoric stresses have a strong influence on the HP structural behavior of scheelite-structured oxides.

Before concluding commenting on the two phases observed below 13.1 GPa, we would like to make some comparison between the Raman spectra of PbWO$_4$-III and scheelite-type PbWO$_4$. In addition to the increase of the number of modes, one of the most distinctive features of the HP phase is the disappearance of the phonon gap typical of scheelite-structured compounds [15, 29]. This observation is in agreement with the fact that in the PbWO$_4$-III structure the tungsten coordination is increased to six (it is four in the scheelite structure), being the WO$_6$ octahedra not isolated as the WO$_4$ tetrahedra are in scheelite-type PbWO$_4$. Consequently, the Raman modes of PbWO$_4$-III cannot be classified as internal and external modes of the oxoanion of tungsten, as is usually done for the scheelite structure [29]. On the other hand, in spite of the coordination change of tungsten induced by the phase transition, the pressure coefficients of the Raman modes are comparable in the low- and high-pressure phases, being the largest one 5 cm$^{-1}$/GPa in scheelite and 4.4 cm$^{-1}$/GPa in PbWO$_4$-III. This fact suggests that both phases are expected to have a similar compressibility, since the pressure derivatives from vibrational frequencies are known to correlate with the bulk modulus in many oxides [35, 36]. Besides, as scheelite, PbWO$_4$-III has modes which gradually soften under compression (two instead of one).

Fig. 4 shows Raman spectra collected from 13.1 GPa up to 47.7 GPa together with two spectra collected upon decompression. In the figure, it can be seen that changes occur in the Raman spectrum at 15.1 GPa indicating the occurrence of a second phase transition to a phase we will name as phase IV. In this phase sixteen Raman modes are identified. The existence of this second transition is consistent with the detection of a phase transition in x-ray absorption experiments at 16.7 GPa [14]. Additional changes occur in the Raman spectrum at 21.2 GPa indicating the onset of a third phase transition to a phase we will label as phase V. This transition occurs together with the visual observation of the formation of multiple domains in the PbWO$_4$ crystal. Note that phases IV and V coexist at 21.2 GPa. Phase V is found as a single phase at 25.5 GPa. For this phase the peaks are broader than for the other three phases. In spite of it, at least twelve modes can be identified. The two most intense of them can be observed under compression up to 43.3 GPa, indicating that phase V is stable up to this pressure. Tables 3 and 4 show the wavenumbers and pressure coefficients of the Raman modes identified for phases IV and V, respectively. Figs. 2 and 3 show the pressure dependence of the Raman modes for phases IV and V. There it can be seen that both transitions are clear not only for the abrupt change in Raman frequencies, but also in their pressure dependences. The most interesting facts to remark is that in phase IV there are modes with very large pressure coefficient, 8.9 cm$^{-1}$/GPa, and two modes with small negative pressure coefficients (see Table 3). On the other hand, phase V has two modes that clearly soften under compression (see Table 4). This phase also has some modes that follow a non-linear behavior (see Fig. 3). It should be noted here that in phase V we observed changes on the intensities of the Raman modes as pressure increased, but no abrupt changes in their pressure evolution. We think the intensities changes can be due to the presence of multiple domains in the crystal, whose number growth with pressure, and which may have different orientations.

At 47.7 GPa no more Raman modes can be detected. This can be caused by a possible pressure-induced amorphization as proposed for BaWO$_4$ at 47 GPa based upon XRD experiments [14] and at 45 GPa for CaWO$_4$, SrWO$_4$, BaWO$_4$, and PbWO$_4$ based upon *ab initio* calculations [1]. According to molecular-dynamic calculations [1] amorphization could be a common phenomenon of scheelite-type compounds. This phenomenon is associated to an abrupt increase of the coordination number of the W atom while the polyhedra around the divalent cation (e.g. Pb) are considerably distorted. Such changes in the coordination polyhedra induce a structural disorder [1] which is compatible with the large decrease of Raman intensity and the appearance of a broad band in our experiments at 47.7 GPa. The presence of the broad band at high frequencies is probably due to the scattering from the continuum density of states arising from the amorphous structure. Unfortunately, the amorphization hypothesis cannot be confirmed only based upon the present Raman measurements. Additional studies are needed to explore the possible pressure-driven amorphization. There is an interesting fact to mention here. The changes observed in the Raman spectra are partially reversible on decompression, showing some hysteresis. Upon pressure release, at 4.6 GPa the Raman spectrum can be identified with phase III. However, the most intense Raman mode of phase I (scheelite) is already present. This is the highest frequency mode of the spectrum and it is denoted by the symbol $ in Fig. 4. Upon further decompression, at 0.1 GPa, a nice Raman spectrum of the scheelite phase is recovered at 0.1 GPa. However, the two most intense modes of phase III are still present, being denoted by asterisks in Fig. 4. This fact is in agreement with the Raman spectrum measured by Christofilos *et al.* upon decompression [16]. However, Manjon *et al.* measured a pure scheelite phase after pressure release. This fact indicates that the reversibility of the phase I to phase III transition could depend on the way pressure is released. On the other hand, the recovery of a crystalline state upon decompression from 47.7 GPa, suggest that if the changes observed at 47.7 are caused by amorphization, then amorphization is reversible, being $PbWO_4$ a phase-change memory material [37] or that the amorphization is not complete.

Regarding the identification of the crystal structure of phases IV and V, it is obvious that it cannot be done based only upon our measurements. Indeed, we expect our results will trigger further *ab initio* calculations and XRD experiments carried out using Ne or He as pressure medium to fully understand the HP structural sequence of $PbWO_4$. Keeping this in mind, we will make here a few comments on the possible structures of phases IV and V. The $HgWO_4$-type [38, 39] and wolframite-type [39, 40] structures have been considered as potential HP structures of $PbWO_4$ [13]. The second one is unlikely to be the post-$PbWO_4$-III structure because such a transition would imply a coordination number decrease for Pb. In addition, the Raman spectra we measured for phase IV are qualitatively different from those measured in different wolframite-type tungstates. In contrast the HgWO$_4$-type structure is built up with similar polyhedral units than PbWO$_4$-III; therefore a transition to this structure is more likely to occur than to wolframite. We have observed that the Raman spectra of phase IV resemble very much those reported for HgWO$_4$ [39]. In particular, the distribution of modes in three isolated frequency regions ($\omega < 200$ cm$^{-1}$, $300$ cm$^{-1} < \omega < 600$ cm$^{-1}$, $\omega > 800$ cm$^{-1}$) is very typical of the HgWO$_4$ structure [39]. Obviously this fact by itself is not sufficient for the identification of the crystal structure of phase IV, but suggests that the HgWO$_4$-type structure should be considered as a possible candidate structure in future XRD experiments and *ab initio* calculations. The last fact we would like to mention is that PbWO$_4$-V becomes stable at pressure where calculations predict that an orthorhombic structure, with space group *Cmca*, is stable [15]. The Raman spectra of phase V resemble those reported for a similar orthorhombic structure (also with space group *Cmca*) in SrMoO$_4$ [34]. Note that in this compound the post-scheelite structure is isomorphic to PbWO$_4$-III. All these facts could be mere coincidences, but they also highlight that such structure should be considered as a possible HP phase in PbWO$_4$.

**4.   Concluding Remarks**

We performed room-temperature Raman scattering measurements under pressure in PbWO$_4$ up to 48 GPa using neon as pressure-transmitting medium. We have observed the occurrence of three phase transitions. The onsets of the transitions are at 6.8, 15.5, and 21.2 GPa. The first transition is from the scheelite structure to the PbWO$_4$-III structure. The crystalline structure of the phases observed beyond 15.5 GPa are not yet identified, but some candidates are suggested based into the comparison with the HP behavior of related compounds. Upon further compression, at 47.7 GPa evidence of a possible pressure-induced amorphization is detected. All structural changes are reversible. The wavenumber and pressure dependence for all the observed phonons are reported. For the scheelite phase the thirteen Raman-active phonons are reported. For the PbWO$_4$-III phase forty-eight of the seventy-two expected modes are reported, which is eleven modes more than previously reported [15]. The results are compared with previous theoretical and experimental studies on PbWO$_4$ [1, 12 – 17, 33]. The reported results show that deviatoric stresses have an important influence on the HP behavior of PbWO$_4$, as happen in isomorphic BaWO$_4$ [20]. In particular, the fergusonite phase observed in several experiments performed using lesser hydrostatic pressure media than Ne is not detected in our present experiments with Ne as the pressure transmitting medium. We hope the new findings reported here will trigger new calculations and experiments to deepen the knowledge of the HP structural behavior of PbWO$_4$.

**Acknowledgments**

D.E. thanks the financial support provided by the Spanish government MINECO under Grant No: MAT2013-46649-C4-1-P. K.K.M. thanks G. Amarendra and B.V.R Tata, MSG, IGCAR for support and encouragement.

**Table 1**: Raman frequencies (ω) and pressure coefficients (dω/dP) for the scheelite structure at ambient conditions. Present results are compared with the literature. The numbers in the parentheses are standard errors to the least significant digit.

|      | This work |      | Theory [11] |      | Experiment [11] |      | Experiment [12] |      | Experiment [9] |      |
|------|-----------|------|-------------|------|-----------------|------|-----------------|------|----------------|------|
| Mode | ω (cm$^{-1}$) | dω/dP (cm$^{-1}$/GPa) | ω (cm$^{-1}$) | dω/dP (cm$^{-1}$/GPa) | ω (cm$^{-1}$) | dω/dP (cm$^{-1}$/GPa) | ω (cm$^{-1}$) | dω/dP (cm$^{-1}$/GPa) | ω (cm$^{-1}$) | dω/dP (cm$^{-1}$/GPa) |
| $B_g$ | 57(1) | -0.8(2) | 52 | 0.7 | 58 | -1.1 | 56 | -0.3 | 59 | -1.0 |
| $E_g$ | 67(1) | 2.2(2) | 64 | 2.3 | 65 | 1.8 | 64 | 3.3 | 67 | 3.4 |
| $B_g$ | 78(1) | 3.6(2) | 79 | 4.4 | 77 | 3.3 | 77 | 3.7 | 80 | 2.8 |
| $E_g$ | 92(1) | 3.1(2) | 92 | 4.6 | 90 | 2.3 | 90 | 3.1 | 93 | 2.5 |
| $A_g$ | 179(1) | 4.0(2) | 191 | 3.3 | 178 | 3.3 | 178 | 4.3 | 182 | 4.0 |
| $E_g$ | 193(1) | 5.0(2) | 193 | 4.6 | 193 | 4.2 | 192 | 5.7 |  |  |
| $A_g$ | 323(1) | 2.2(2) | 310 | 2.1 | 323 | 1.9 | 323 | 1.8 |  |  |
| $B_g$ | 328(1) | 2.3(2) | 311 | 3.0 | 328 | 2.1 | 328 | 2.2 | 329 | 3.1 |
| $B_g$ | 357(1) | 3.2(2) | 350 | 2.7 | 357 | 2.8 | 356 | 3.6 |  |  |
| $E_g$ | 360(1) | 3.2(2) | 351 | 2.8 | 362 | 2.7 |  |  | 360 | 3.4 |
| $E_g$ | 752(1) | 2.4(2) | 750 | 3.1 | 752 | 2.4 | 752 | 2.7 | 755 | 3.4 |
| $B_g$ | 766(1) | 0.8(2) | 758 | 1.7 | 766 | 0.9 | 765 | 1.0 | 769 | 0.6 |
| $A_g$ | 906(1) | 1.3(2) | 890 | 1.4 | 906 | 0.8 | 905 | 1.4 | 908 | 1.2 |

**Table 2**: Raman frequencies of PbWO$_4$-III at different pressures compared with previous experiments [15, 33] and calculations [15]. Pressure coefficients from the present work are also included. The numbers in the parentheses are standard errors to the least significant digit.

| Mode | Theory Ref. 15 8.7 GPa | Experiment This work 8.9 GPa | | Experiment Ref. 33 8.9 GPa | Experiment Ref. 15 8.9 GPa | Experiment This work 13.1 GPa | Experiment Ref. 15 13.7 GPa |
|---|---|---|---|---|---|---|---|
| | $\omega$ (cm$^{-1}$) | $\omega$ (cm$^{-1}$) | d$\omega$/dP (cm$^{-1}$/GPa) | $\omega$ (cm$^{-1}$) | $\omega$ (cm$^{-1}$) | $\omega$ (cm$^{-1}$) | $\omega$ (cm$^{-1}$) |
| B$_g$ | 43 | | | | | | |
| A$_g$ | 44 | 45(1) | 1.0(2) | | | 50(1) | 46(1) |
| B$_g$ | 52 | | | | | | |
| A$_g$ | 56 | 55(1) | 0.2(2) | | 60(2) | 56(1) | 61(1) |
| A$_g$ | 59 | | | | | | |
| A$_g$ | 63 | | | | | | |
| B$_g$ | 66 | 66(1) | 0.0(2) | | 68(2) | 66(1) | 68(1) |
| B$_g$ | 69 | | | | 63(2) | | 69(1) |
| A$_g$ | 72 | 72(1) | 0.4(2) | | 70(2) | 75(1) | 73(1) |
| A$_g$ | 74 | | | | | | |
| B$_g$ | 81 | 80(1) | 0.2(2) | | 84(2) | 81(1) | 84(1) |
| A$_g$ | 87 | 88(1) | 1.2(2) | | | 94(1) | |
| B$_g$ | 91 | | | | | | |
| B$_g$ | 101 | | | | | | |
| A$_g$ | 104 | 107(1) | 1.0(2) | | 117(2) | 112(1) | 120(1) |
| B$_g$ | 109 | | | | 123(2) | | 123(1) |
| A$_g$ | 118 | 120(1) | 0.8(2) | | 127(3) | 124(1) | 130(2) |
| B$_g$ | 121 | 125(1) | 3.4(2) | | 123(3) | 147(1) | 145(2) |
| A$_g$ | 130 | | | | | | |
| B$_g$ | 136 | 137(1) | 0.2(2) | | | 139(1) | |
| B$_g$ | 142 | 140(1) | -0.6(2) | | | 137(1) | |
| A$_g$ | 153 | 155(1) | 3.6(2) | | | 173(1) | |
| B$_g$ | 170 | 165(1) | 2.5(2) | 167(1) | | 173(1) | |
| A$_g$ | 171 | | | | | | |
| A$_g$ | 173 | 174(1) | 3.0(2) | | | 184(1) | |
| B$_g$ | 179 | | | 181(1) | | | |
| A$_g$ | 191 | 186(1) | 0.8(2) | 182(1) | 185(2) | 190(1) | 189(1) |
| B$_g$ | 196 | 196(1) | 1.2(2) | 194(1) | 195(2) | 202(1) | 201(1) |
| B$_g$ | 213 | 206(1) | 1.4(2) | 209(1) | 202(2) | 213(1) | 208(1) |
| A$_g$ | 215 | 214(1) | 0.8(2) | | | 218(1) | |
| B$_g$ | 232 | 225(1) | 2.5(2) | 229(1) | 221(2) | 233(1) | 222(5) |
| B$_g$ | 237 | | | | | | |
| A$_g$ | 244 | 241(1) | 2.4(2) | | 240(2) | 253(1) | 246(1) |
| A$_g$ | 255 | | | | | | |

**Table 2**: Continuation

| | | | | | | | |
|---|---|---|---|---|---|---|---|
| $B_g$ | 258 | | | | | | |
| $A_g$ | 259 | 264(1) | 3.6(2) | 259(1) | | 282(1) | |
| $A_g$ | 269 | | | 275(1) | | | |
| $B_g$ | 283 | 292(1) | -2.2(2) | 285(1) | 277(2) | 283(1) | 266(1) |
| | | | | | 288(2) | | 288(1) |
| $B_g$ | 306 | 306(1) | 0.1(2) | 299(1) | 305(2) | 307(1) | 305(1) |
| $A_g$ | 309 | | | | | | |
| $A_g$ | 321 | 324(1) | 2.4(2) | 325(1) | | 336(1) | |
| $B_g$ | 332 | | | | | | |
| $A_g$ | 340 | 342(1) | 1.0(2) | 339(1) | | 346(1) | |
| $B_g$ | 341 | | | | | | |
| $B_g$ | 362 | 354(1) | 1.8(2) | 357(1) | | 363(1) | |
| $A_g$ | 368 | | | | | | |
| $A_g$ | 380 | | | 385(1) | | | |
| $B_g$ | 390 | 391(1) | 0.2(2) | 387(1) | 390(2) | 392(1) | 390(1) |
| $A_g$ | 391 | 397(1) | 1.0(2) | 396(1) | | 402(1) | 400(1) |
| $B_g$ | 400 | 408(1) | 2.0(2) | 414(1) | 400(3) | 415(1) | 417(2) |
| $B_g$ | 436 | 423(1) | 1.4(2) | 419(1) | 419(2) | 430(1) | 430(1) |
| $A_g$ | 438 | | | | | | |
| $A_g$ | 467 | 451(1) | 3.4(2) | 462(1) | | 468(1) | |
| $B_g$ | 472 | 480(1) | 1.2(2) | 488(1) | 485(3) | 486(1) | 490((3) |
| $A_g$ | 495 | 498(1) | 0.8(2) | | 496(3) | 508(1) | 501(2) |
| $B_g$ | 503 | 517(1) | 4.4(2) | 521(1) | 520(3) | 532(1) | 542(2) |
| $A_g$ | 575 | 548(1) | 3.4(2) | | 539(4) | 565(1) | 564(3) |
| $B_g$ | 576 | 594(1) | 2.4(2) | 595(1) | 597(4) | 606(1) | 605(3) |
| $A_g$ | 644 | | | | | | |
| $B_g$ | 645 | 648(1) | 2.0(2) | | 671(4) | 658(1) | 672(3) |
| $A_g$ | 694 | 672(1) | 1.0(2) | 662(1) | 690(4) | 674(1) | 693(3) |
| $B_g$ | 697 | 698(1) | 2.4(2) | 704(1) | 705(3) | 710(1) | 710(2) |
| $B_g$ | 718 | | | | | | |
| $A_g$ | 719 | | | 743(1) | | | |
| $A_g$ | 730 | 730(1) | 3.4(2) | 748(1) | | 747(1) | |
| $B_g$ | 761 | 772(1) | 1.6(2) | 746(1) | 765(3) | 781(1) | 773(2) |
| $A_g$ | 774 | 793(1) | 1.6(2) | 797(1) | 792(3) | 801(1) | 799(2) |
| $B_g$ | 802 | 813(1) | 2.0(2) | 826(1) | 816(4) | 823(1) | 825(3) |
| $A_g$ | 857 | 873(1) | 1.0(2) | | | 878(1) | |
| $B_g$ | 869 | 895(1) | 1.4(2) | 887(1) | 891(2) | 902(1) | 899(1) |
| $A_g$ | 912 | 905(1) | 2.2(2) | 916(1) | 902(2) | 916(1) | 914(1) |
| | | 931(1) | 0.8(2) | | 931(3) | 935(1) | 935(2) |
| $B_g$ | 919 | 945(1) | 1.4(2) | | 942(3) | 952(1) | 949(2) |

Table 3: Raman frequencies and pressure coefficients of PbWO$_4$-IV at 15.5 GPa. The numbers in the parentheses are standard errors to the least significant digit.

| $\omega$ (cm$^{-1}$) | d$\omega$/dP (cm$^{-1}$/GPa) | $\omega$ (cm$^{-1}$) | d$\omega$/dP (cm$^{-1}$/GPa) |
|---|---|---|---|
| 53(2) | 0.8(4) | 309(2) | -0.2(4) |
| 66(2) | 1.0(4) | 371(2) | 2.4(4) |
| 77(2) | 0.6(4) | 440(2) | 0.6(4) |
| 96(2) | 1.4(4) | 463(2) | 8.9(4) |
| 125(2) | 0.4(4) | 580(2) | 5.2(4) |
| 153(2) | -0.2(4) | 826(2) | 4.1(4) |
| 177(2) | 0.7(4) | 886(2) | 3.5(4) |
| 194(2) | 1.1(4) | 910(2) | 2.2(4) |

Table 4: Raman frequencies and pressure coefficients of PbWO$_4$-V at 29.7 GPa. The numbers in the parentheses are standard errors to the least significant digit.

| $\omega$ (cm$^{-1}$) | d$\omega$/dP (cm$^{-1}$/GPa) | $\omega$ (cm$^{-1}$) | d$\omega$/dP (cm$^{-1}$/GPa) |
|---|---|---|---|
| 61(2) | 0.0(4) | 451(2) | 2.9(4) |
| 83(2) | 0.7(4) | 638(2) | 4.1(4) |
| 90(2) | 1.0(4) | 769(2) | 2.6(4) |
| 162(2) | 1.1(4) | 851(2) | -3.5(4) |
| 373(2) | -5.4(4) | 932(2) | 1.5(4) |
| 482(2) | 2.1(4) | 962(2) | 2.5(4) |

**Figure captions to be written**

**Figure 1:** (color online) Raman spectra measured in PbWO$_4$ up to 13.1 GPa. The onset of the scheelite-PbWO$_4$-III transition takes place at 6.8 GPa. Pressures in different phases are indicated in the figure. Ticks are used to show Raman modes of phases I and III.

**Figure 2:** Pressure dependence of the Raman modes (low-frequency region) for different phases of PbWO$_4$. Empty and solid symbols are used to facilitate the identification of different modes. Circles: phase I, squares: phase III, diamonds: phase IV, and hexagons: phase V. The solid lines are guides to the eye.

**Figure 3:** Pressure dependence of the Raman modes (high-frequency region) for different phases of PbWO$_4$. Empty and solid symbols are used to facilitate the identification of different modes. Circles: phase I, squares: phase III, diamonds: phase IV, and hexagons: phase V. The solid lines are guides to the eye.

**Figure 4:** (color online) Raman spectra measured in PbWO$_4$ from 13.1 GPa to 47.7 GPa. The onset of the phase transitions take place at 15.5 and 21.2 GPa. Amorphization is detected at 47.7 GPa. Two Raman spectra measured upon decompression are shown to illustrate the reversibility of the structural changes. Pressures are different phases are indicated in the figure. Ticks identify the Raman modes of the recovered scheelite-type phase.

Figure 1

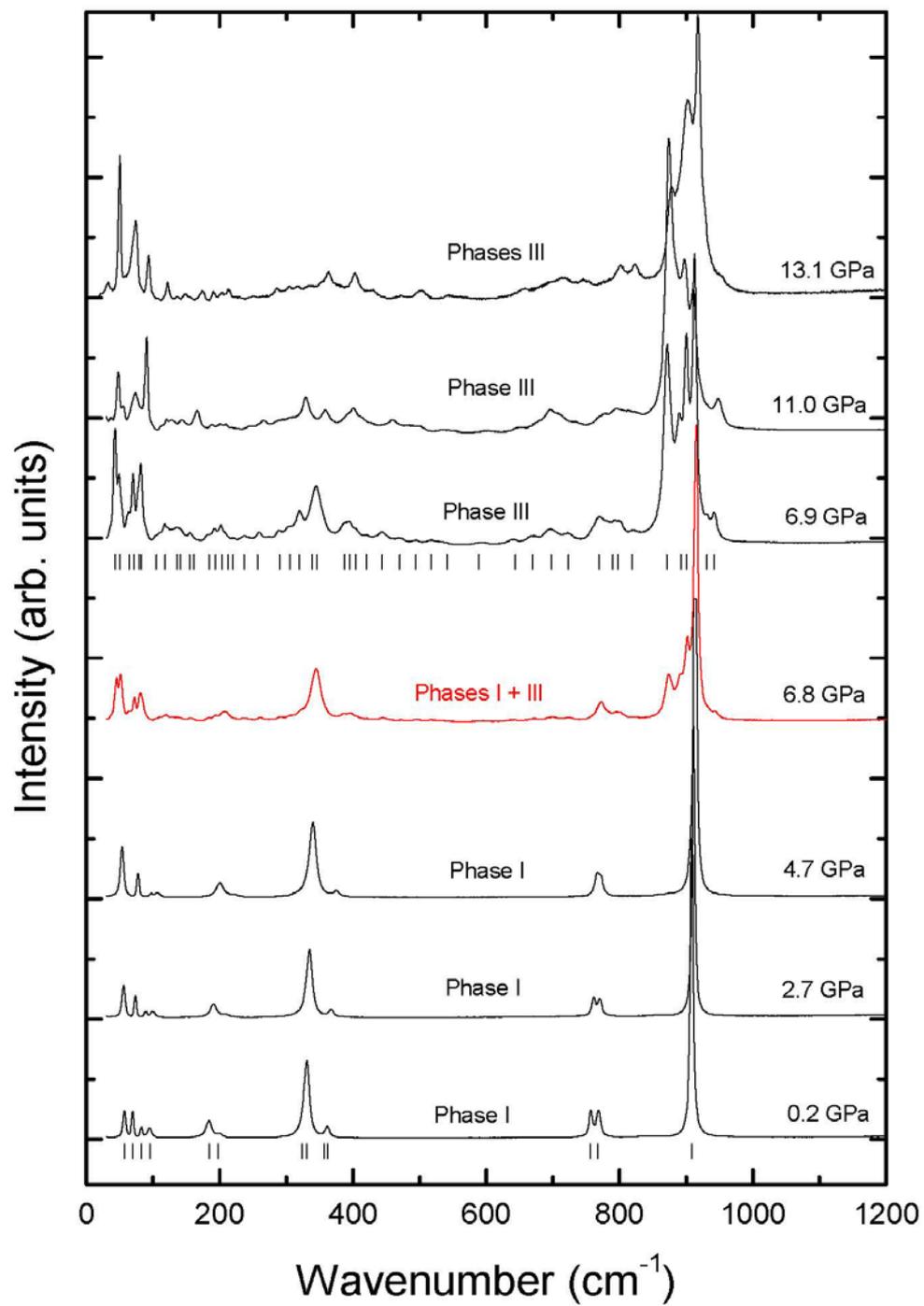

Figure 2

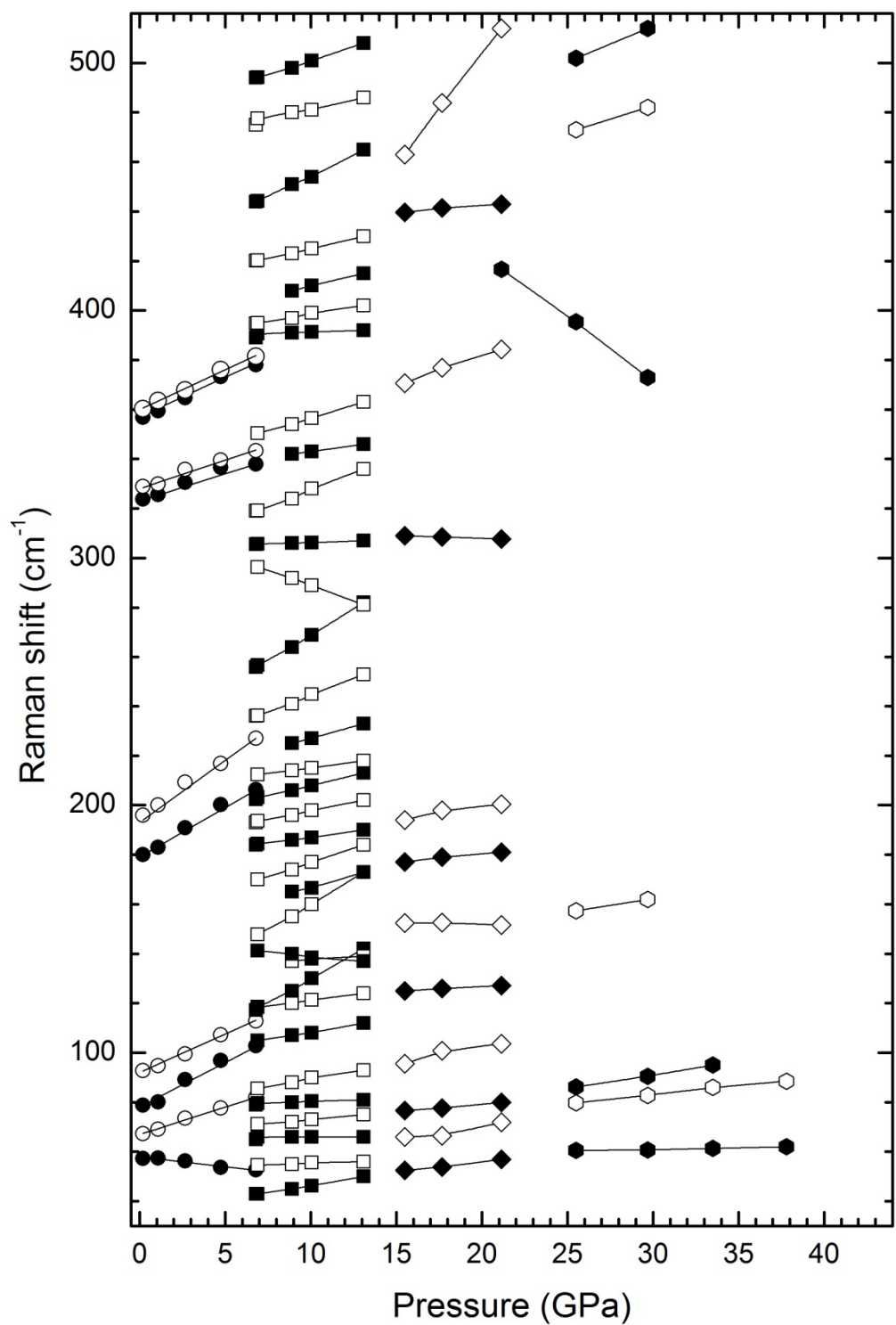

Figure 3

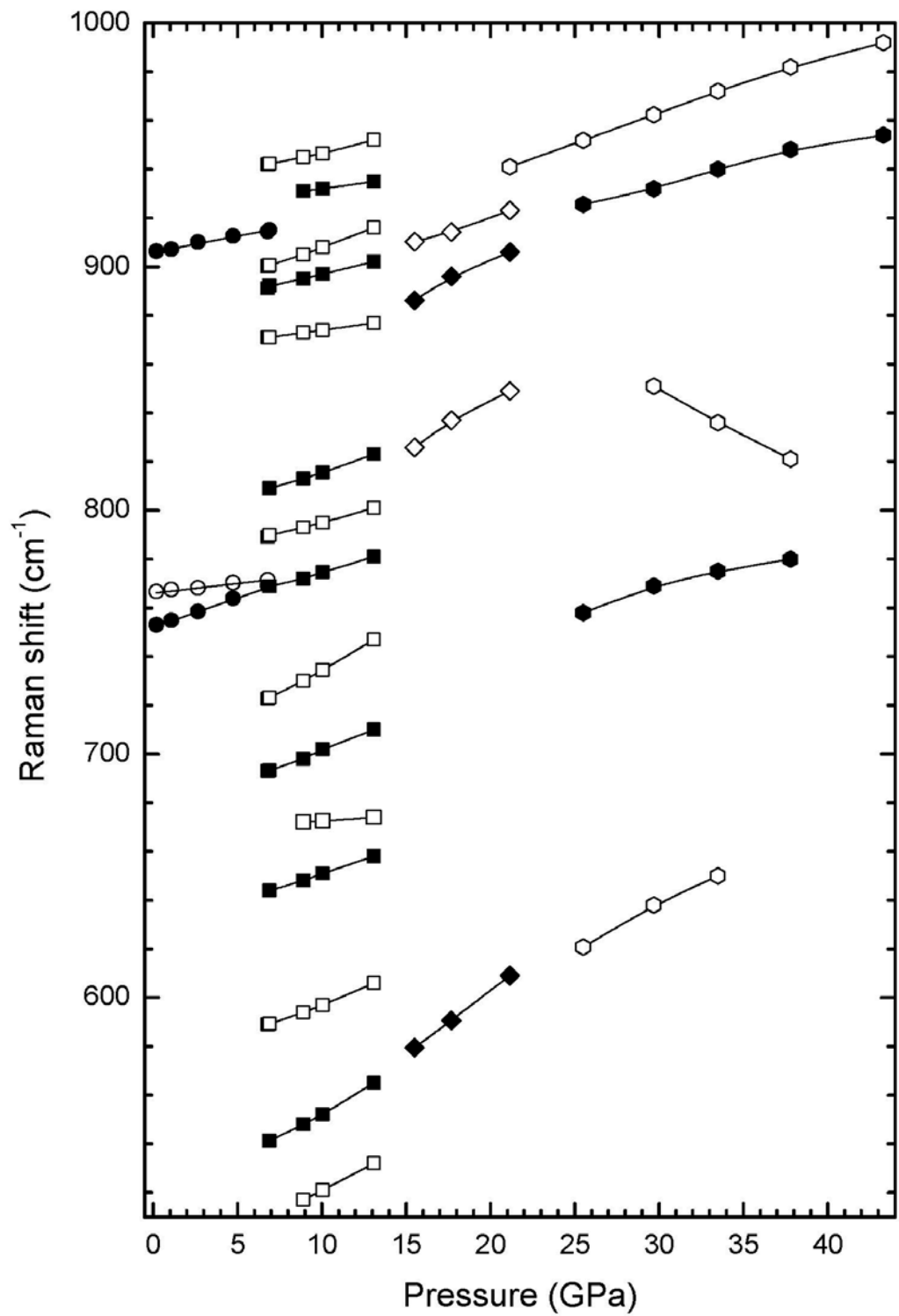

Figure 4

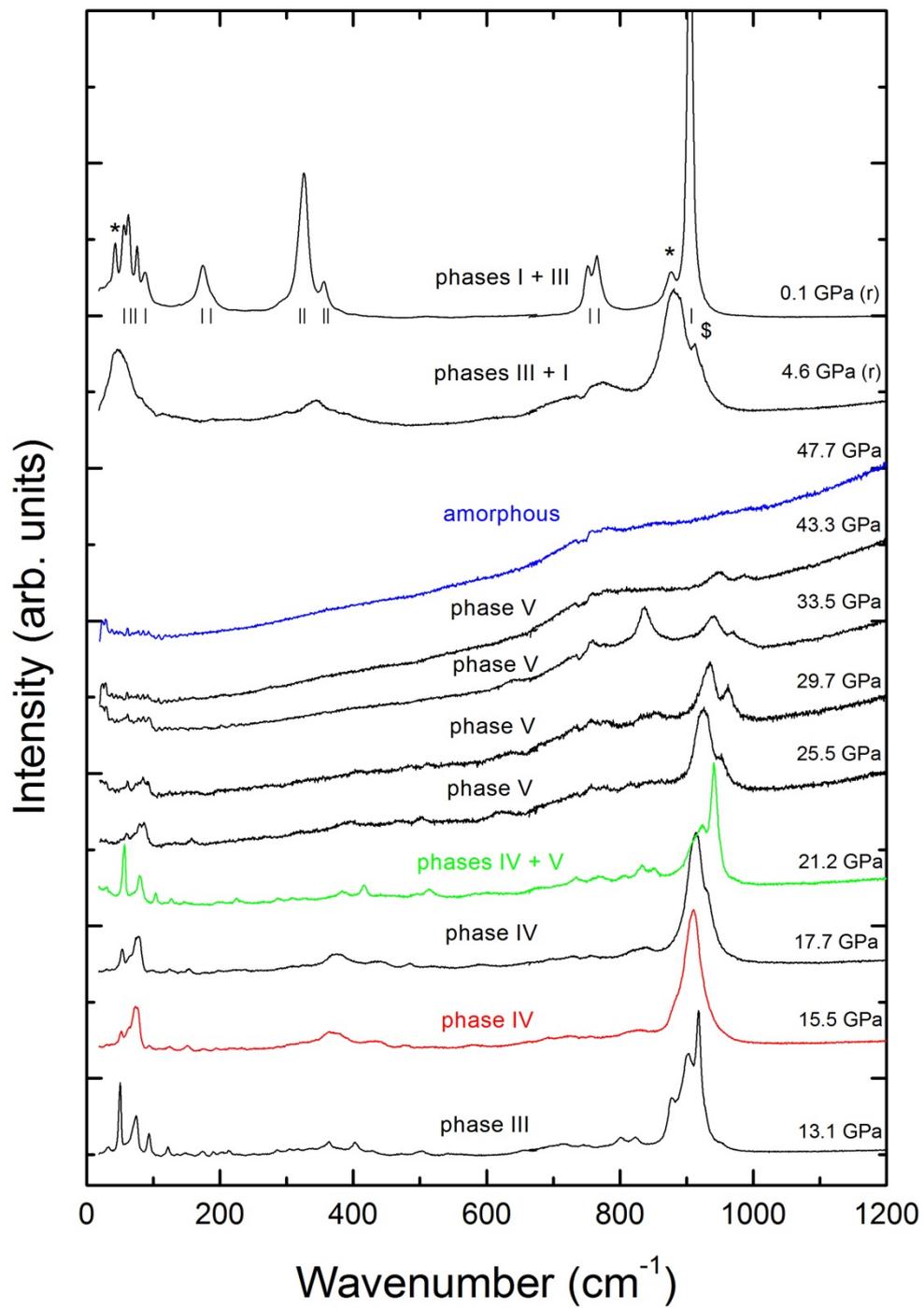